\documentclass[]{emulateapj}

\def\bibfiles{biblio}
\def\aareferences{\bibliographystyle{apj}
                  \bibliography{aajour,\bibfiles}}

\def\rmit#1{{\it #1}}              

\def\eg{\rmit{e.g.}}

\usepackage{natbib,graphicx}

\usepackage{color}

\shorttitle{Scattering of $f-$modes by small magnetic elements}

\shortauthors{Felipe et al.}

\begin{document}

\title{Scattering of the $f-$mode by small magnetic flux elements from observations and numerical simulations}

\author{T. Felipe\altaffilmark{1}, D. Braun\altaffilmark{1}, A. Crouch\altaffilmark{1}, A. Birch\altaffilmark{1,2}}
\email{tobias@cora.nwra.com}

\altaffiltext{1}{NorthWest Research Associates, Colorado Research Associates, Boulder, CO 80301, USA}
\altaffiltext{2}{Max-Planck-Institut f\"{u}r Sonnensystemforschung, Max-Planck-Str. 2, 37191 Katlenburg-Lindau, Germany}

\begin{abstract}
The scattering of $f-$modes by magnetic tubes is analyzed using three-dimensional numerical simulations. An $f-$mode wave packet is propagated through a solar atmosphere embedded with three different flux tube models which differ in radius and total magnetic flux. A quiet Sun simulation without a tube present is also performed as a reference. Waves are excited inside the flux tube and propagate along the field lines, and jacket modes are generated in the surroundings of the flux tube, carrying $40\%$ as much energy as the tube modes. The resulting scattered wave is mainly an $f-$mode composed of a mixture of $m=0$ and $m=\pm 1$ modes. The amplitude of the scattered wave approximately scales with the magnetic flux. A small amount of power is scattered into the $p_1-$mode. We have evaluated the absorption and phase shift from a Fourier-Hankel decomposition of the photospheric vertical velocities. They are compared with the results obtained from the emsemble average of 3400 small magnetic elements observed in high-resolution MDI Doppler datacubes. The comparison shows that the observed dependence of the phase shift with wavenumber can be matched reasonably well with the simulated flux tube model. The observed variation of the phase-shifts with the azimuthal order $m$ appears to depend on details of the ensemble averaging, including possible motions of the magnetic elements and asymmetrically shaped elements.

\end{abstract}



\section{Introduction}

The work by \citet{Braun+etal1988} has shown that sunspots can absorb up to half the power of incident $p-$modes. At the same time, part of the wave flux is scattered. The $p-$mode absorption and scattering phase shifts depend on the frequency, degree, radial order, and azimuthal order of the incident mode \citep{Bogdan+etal1993,Braun1995}, and also on the magnetic structure, making the study of the scattering a promising way to infer the subsurface structure of sunspots and other magnetic features. In these studies the authors used Hankel analysis, a method which decomposes the $p-$modes into inward and outward propagating waves in annuli surrounding the sunspot.

Several mechanisms have been proposed to explain the observed absorption, mode conversion
\citep{Cally+Bogdan1993} being the principal candidate. When an acoustic wave encounters a magnetic field concentration, it is split into fast and slow modes. This mode transformation occurs at the height where the Alfv\'en and sound velocities are comparable, since at that layer the distinction between the modes is small. Below this height the sound speed is higher than the Alfv\'en speed and the modes are effectively decoupled. The fast mode is an acoustic-like wave, while the slow mode is similar to an Alfv\'en wave and propagates along field lines removing energy from the acoustic wave. The observed absorption of $f-$modes can be accounted for by a vertical magnetic field \citep{Cally+etal1994}, but the $p-$mode absorption obtained from this model is insufficient. However, the presence of inclined magnetic field produces significant increases in absorption with a peak at around 30$^o$ \citep{Crouch+Cally2003}, which are consistent with observed values \citep{Cally+etal2003}. Other mechanisms may also play a role. One of them is resonant absorption \citep{Hollweg1988, Rosenthal1992}. It may occur when the flux tube has a smooth variation of the magnetic field rather than discontinuous, and represents absorption of wave energy by the transition layer when the incident acoustic waves resonantly excite MHD waves in the magnetic structure. However, the amount of absorption achieved by this mechanism cannot explain the observed loss of $p-$mode power. \citet{D'Silva1994} points out that apart from this absorption produced by the dissipation of the $p-$modes in resonant layers and mode conversion, mode mixing also takes places. In this process the power of an incident $p-$mode mode with a certain frequency and degree $l$ can be dispersed into an outgoing $p-$mode with the same frequency, but different degree.

Flux tubes are a key feature to understand solar magnetic activity. They are spread all over the solar surface and couple different layers of the solar atmosphere. It has been proposed that magnetic flux tubes can act as wave guides, being one of the possible sources which supply energy to the upper layers to account for the chromospheric and coronal heating \citep{Jefferies+etal2006}. The interaction of $p-$modes with thin flux tubes excites tube waves, including sausage waves, which are axisymmetric, longitudinal waves driven by variations in the total pressure, and kink waves, whose restoring force is magnetic tension and buoyancy, producing transversal oscillations. These waves propagate upward or downward and extract energy from the $p-$modes of the acoustic cavity \citep{Bogdan+etal1996, Hindman+Jain2008, Jain+etal2009}. The kink mode is driven by the distortion of the tube produced by the harmonic flow field of the $p-$modes, while the sausage wave is excited by the pressure pertubations associated to the acoustic waves \citep{Bogdan+etal1996}. These mechanisms are different from mode conversion, discussed in the previous paragraph, where fast and slow magnetoacoustic waves exchange energy due to their strong coupling in the region where the sound and Alfv\'en speeds are comparable. Recently, \citet{Daiffallah+etal2011} used numerical simulations to study the scattering of an $f-$mode by vertical flux tubes of different sizes, finding that the scattering by tubes with small radius is dominated by the kink mode, while the sausage mode is dominant for large tubes. This result coincides with the earlier work by \citet{Bogdan+etal1996}, who studied analytically the nature of wave interactions with thin flux tubes \citep{Spruit1981} and found that the kink mode is the dominant tube wave. The thin flux tube approximation assumes that the diameter of the tube is smaller in comparison to the pressure scale height, and thus the horizontal variations inside it can be neglected. \citet{Hanasoge+etal2008} evaluated the scattering matrix associated with an $f-$mode that interacts with a thin flux tube in a stratified atmosphere, focusing on the kink mode excited in the magnetic tube. They found that most of the scattered wave corresponds to an $f-$mode with amplitude of 1.17\% and with a phase shift of abound 50$^o$ relative to the incident wave, overstimating the observed value by a factor of 8.8 \citep{Duvall+etal2006}. On the other hand, the recent work by \citet{Hindman+Jain2011} analyzed the axisymmetric scattering of $p-$modes, mediated through the excitation of sausage waves on the flux tube, instead of the kink mode. They obtained a small absorption due to the poor coupling between the $f-$mode and the sausage mode for thin flux tubes, as pointed out by some of the works previously described in this paragraph.

Although these theoretical works have provided the first predictions about the modification of the solar wave field produced by flux tubes, as far as our concern no attempt has been made to observationally measure the detailed properties of the scattering produced by these small magnetic elements, with the exception of the estimates of amplitude and phase of monopole and dipole scattering by \citet{Duvall+etal2006}. One of the objectives of this work is to present the measurement of the phase shift and its variations with the azimuthal orders $m$ and degree $L$. This data is a fundamental input to confront with the theory. On the other hand, most theoretical studies of this topic have been based on an analytical development. All of them have been restricted by some limitations, including the use of the thin flux tube approximation, the lack of the gravitational stratification, the analysis of a polytrope instead of a realistic solar atmosphere, or some constrains in the process that mediates the scattering. Numerical simulations are a more versatile approach and allow us to study more general situations.

In this work we study the scattering of an $f-$mode by flux tubes of different radius and magnetic flux using numerical simulations. As discussed previously, it is well known that thin flux tubes support sausage and kink modes \citep{Roberts+Webb1978, Spruit1981}. On the other hand, in unstratified atmospheres permeated by homogeneous magnetic fields one would expect the propagation of pure fast and slow magnetoacoustic waves and Alfv\'en mode. In atmosphere stratified by gravity (for example) the fast, slow, and Alfv\'en waves are coupled in general and this distinction between modes no longer applies, but even in these cases it is useful to refer to this simple picture to discuss the properties of fast and slow magnetoacoustic-like waves (in regions where the sound and Alfv\'en speed differ greatly). In the case of thick flux tube models, representative of a sunspot, for example, acoustic waves can be converted into these modes by means of mode conversion. The flux tube models presented in this paper correspond to an intermediate case between these two extremes. We expect a smooth transition (with increasing radius) from mostly excitation of the kink and sausage modes at small radius, to excitation of waves that look like the fast and slow magnetoacoustic waves of a thick flux tube (at large radius). However, in this study we made no attempt to distinguish between neither the different wave modes which are present nor the mechanisms that generates them. Instead, we will refer to ``tube modes'' or ``tube waves excitation'' indistinctly. We aim to carry out a direct comparison between the numerical and observational results by performing a Hankel analysis of the data obtained from the interaction of an $f-$mode with flux tubes. The organization of the paper is as follows. In Section \ref{sect:observations} we introduce the observations used in this study. Section \ref{sect:procedures} briefly describes the numerical code and the set up of the simulations. In Sections \ref{sect:scattering} and \ref{sect:jacket}, we present the tube mode excitation, the scattering, and the jacket modes produced in these simulations. The results of the Hankel analysis are shown in Section \ref{sect:hankel}, including a comparison with observations, and finally we conclude with a summary of our calculations and a discussion of their applicability to understanding observations.

\section{Observations}
\label{sect:observations}

The motivation of this paper is to understand
how small magnetic elements affect $f$-mode wave propagation.
The comparison of observations of magnetic elements on the Sun 
with equivalent measurements obtained from numerical simulations
can help infer the properties of the solar magnetic elements. These
comparisons can also be used to assess the limitations of, and potentially improve, the
observational techniques.

Observations of solar magnetic elements, using Dopplergrams obtained from the
Michelson Doppler Imager
\citep[MDI;][]{Scherrer+etal1995} onboard the {\it Solar and
Heliospheric Observatory} (SOHO), were analyzed using the Fourier-Hankel
spectral decomposition method, as described in detail by \cite{Braun1995}.
The goal of the analysis is the decomposition of the observed line-of-sight velocities
into inward and outward propagating waves in an annular region centered on the
flux tubes. This allows us to detect the effect of a magnetic feature
on the wave field through the difference between the outward and inward
radially propagating Hankel components. In a spherical polar coordinate system 
$(\theta, \phi)$ the wave components take the form
\begin{eqnarray}
\label{eq:decomp}
\lefteqn{\Psi _m(\theta ,\phi,t)=e^{i(m\phi +2\pi \nu t)}\times} \nonumber\\
&&\times[A_m(L,\nu)H_m^{(1)}(L\theta )+B_m(L,\nu )H_m^{(2)}(L\theta )],
\end{eqnarray}
where $m$ is the polar azimuthal order, $H_m^{(1)}$ and $H_m^{(2)}$ are Hankel
functions of the first and second kind respectively, $t$ is time, $\nu $ is
the temporal frequency, $L\equiv [l    (l    +1)]^{1/2}$ where $l$ is the spherical
harmonic degree of the mode, and $A_m$ and $B_m$ 
are the complex amplitudes of incoming and outgoing waves respectively.
For small fields of view, such as the data we consider in this paper, the expansion is
well approximated in terms of Hankel functions as opposed to Legendre functions. 

The Hankel analysis was centered on the locations of small magnetic elements 
as well as a larger set of control locations offset by fixed distances from each
magnetic element. There were no active regions present in the MDI observations
used. We hereafter refer to the control locations as ``quiet-Sun'' regions, 
although no attempt was made to select (or exclude) these locations based on 
magnetic properties of the MDI pixels. Thus the primary difference between the 
``magnetic feature'' and ``quiet Sun'' locations is that the former 
coincide with the peaks of small magnetic flux regions while the latter are pseudo-random
locations.
The data consist of the set of 102 ``high-resolution'' 
MDI Doppler and magnetogram datacubes, each with a 4 hr interval,
which has been previously used by \cite{Duvall+etal2006} to measure travel-time kernels 
for time-distance helioseismology. 
Each of the 102 regions are confined to the MDI ``high resolution field'' which
spans 11 $\times$ 11 arc minutes and is centered about 160 arc seconds north of disk center
\citep{Scherrer+etal1995}. The data are obtained from all of the MDI
full-resolution observations of at least 4 hours in duration and spanning the years 1996 and 1997.
The cadence of both the Dopplergrams used in the helioseismic
analysis, and the magnetograms used for identifying small magnetic elements,
is 60 seconds.

\begin{figure}[!ht] 
 \centering
 \includegraphics[width=9cm]{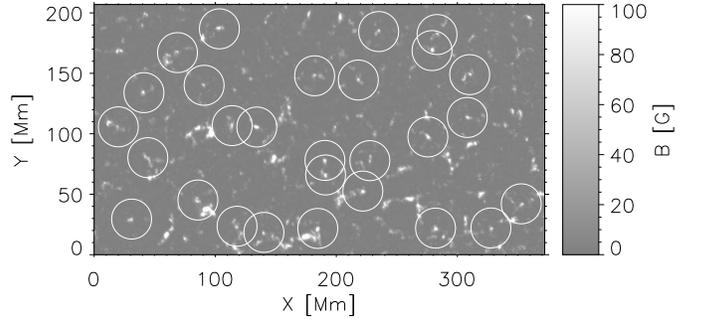}
  \caption{The absolute value of the magnetic flux density, 
  after averaging over 4 hours, for one of the 
  102 MDI regions. White circles indicate the position of the magnetic 
  elements used for the Hankel analysis and their size corresponds to the outer annulus.}
  \label{fig:magnetogram}
\end{figure}

The criteria and procedure for the selection of the magnetic
elements are described in detail by \cite{Duvall+etal2006}. The average magnetic feature
has a peak magnetic flux density of 76 G, and a full-width-at-half-maximum (FWHM)
of 2.6 Mm, as determined from a fit by a 2D Gaussian function to the averaged 
magnetogram \citep{Duvall+etal2006}. 
We used locations of the features from tables provided to us by Duvall. 
In total, nearly 3400 locations of features were used. This is a somewhat larger sample than 
the (approximately) 2500 features used by \cite{Duvall+etal2006} since we used features closer
(i.e.\ as close as 40 pixels) to the edge of the datacubes than used in that work. 
There were four 
quiet-Sun locations offset from each magnetic feature used, thus yielding a quiet-Sun
control sample of 13,600. Figure \ref{fig:magnetogram} shows
the absolute value of the time-averaged magnetic flux density corresponding to
one of the Doppler datacubes, showing the sample of features for that datacube.

The Fourier-Hankel decomposition was performed for all locations, and the coefficients
$A_m (L, \nu)$ and $B_m (L, \nu)$ 
were determined for waves within an annular domain with colatitude relative to the center point of the analysis
$\theta$ ranging from $\theta_{min} = 10$ pixels to $\theta_{min} = 40$ 
pixels, where a pixel corresponds to 0.034 heliocentric degrees (or 0.413 Mm). 
Details of the decomposition method are described by \cite{Braun1995}. The method
consists of discrete numerical transforms in the azimuthal, colatitude and temporal
domains. The azimuthal transform is computed for integer values of azimuthal order
-10 $\leq m \leq$ 10 at the highest values of $L$. Not all of these orders $m$ are
computed (or useful) for smaller values of wavenumbers \citep[see][for details]{Braun1995}.
The values of $L$ and $\nu$ for which the relative colatitude and temporal transforms are computed
compose a grid with spacing $\Delta L = 2 \pi / (\theta_{max} - \theta_{min}) = 352.9$
and $\Delta \nu = 1/T = 0.0694$ mHz, where $T$ is the duration of the observations (4 hr). 

We focus in this paper on comparisons between observations and models of 
phase-shifts between the outgoing and incoming wave components. Our primary 
motivation for this is that the measurement and interpretation of
observations of amplitude differences (e.g.\ due to absorption) are affected
by factors such as the presence of background convective (or instrumental) noise
and by details of the solar excitation and damping mechanisms of
the waves. We make no attempt to realistically include these in the numerical
models described later. This challenges our ability to make meaningful comparisons
of absorption coefficients, for example.
Some of these effects further restrict the measurement of phase-shifts to those
modes (typically with lower wavenumbers and temporal frequencies) 
which have lifetimes significantly longer than the time the waves take
to propagate across the entire annulus \citep{Braun1995}.

To determine feature-averaged phase shifts, we 
compute the summations, over the magnetic element and quiet Sun ensembles, 
of the product $B_m (L, \nu) A^*_m (L, \nu)$ where the asterisk denotes the 
complex conjugate.  The effects of a ``spurious phase-shift' caused by the leakage
of wave amplitudes across the wavenumber-frequency domain \citep[see][]{Braun1995} are
removed by considering the relative {\it difference} in the phase-shift 
between the magnetic-element and quiet-Sun ensembles. Thus, we compute a ``corrected''
ensemble-averaged product, 
\begin{equation}
\langle B_m A^{*}_m {\rangle}^{\prime}_{\rm me} =   | \langle B_m A^{*}_m {\rangle}_{\rm me} |
e^{i [\arg(\langle B_m A^{*}_m {\rangle}_{\rm me} ) - \arg( \langle B_m A^{*}_m {\rangle}_{\rm qs} )]},
\end{equation}
where the brackets indicate the ensemble average over the samples of magnetic elements
(me) and quiet-Sun (qs) locations and the explicit dependencies
of $A_m$ and $B_m$ on $L$ and $\nu$ are omitted for clarity. 

\begin{figure}[!ht] 
 \centering
 \includegraphics[width=9cm]{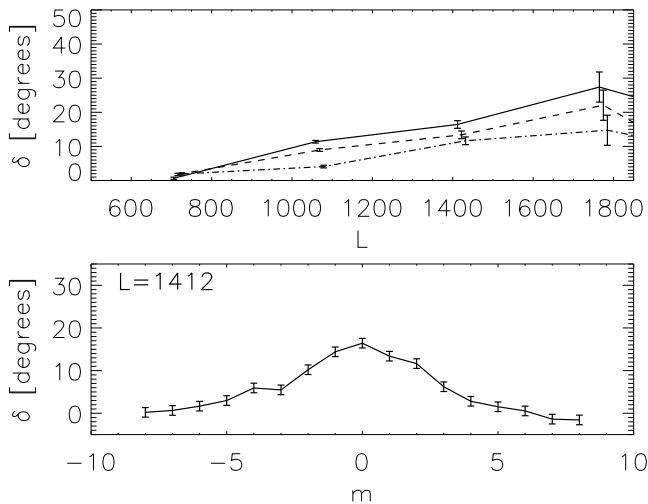}
  \caption{Top panel: Variation of the observational phase shift with L for the azimuthal order 
  $m=0$ (solid line), $m=1$ (dashed line), and $m=2$ (dashed-dotted line). 
  Bottom panel: Variation of the observational phase shift with $m$ at $L=1412$. 
  The error estimates are obtained from an average over $m$ of the absolute value of the 
  phase-shift difference between $+m$ and $-m$, divided by $\sqrt2$. 
  This estimation assumes that the error does not depend on $m$.}
  \label{fig:phase_obs_solo}
\end{figure}

An additional averaging across the width of the $f$-mode ridge, at each wavenumber
$L$, is performed, such that the resulting phase shift is given by
\begin{equation}
{\delta}_{\rm obs} (L) = \arg \left( \int_{\nu_0(L) - \delta \nu}^{\nu_0(L) + \delta \nu} 
\langle B_m (L, \nu) A^{*}_m (L, \nu) {\rangle}^{\prime}_{\rm me} d\nu \right)
\label{eq:phase}
\end{equation}
where $\nu_0(L)$ is the $f$-mode frequency at wavenumber $L$.
The window $\delta \nu$ is determined by 
an inspection of the power spectra $|A_m(L, \nu) |^2$ such that it contains 
most of the $f$-mode ridge power. The value of $\delta \nu$ increases from 0.3 mHz at $L = 706$
to 0.7 mHz at $L = 1765$. 
Figure \ref{fig:phase_obs_solo} shows the observational phase shift, including its variation with $L$ for
several values of $m$, and its variation with $m$ at $L = 1412$. In the next section, we describe the model we construct to reproduce
these phase shifts.

\section{Numerical procedures}
\label{sect:procedures}

We have solved numerically the nonlinear three-dimensional (3D) MHD equations using the code Mancha \citep{Khomenko+Collados2006,Felipe+etal2010a}. The code solves the equations for perturbations, which are obtained by subtracting the equations of initial magnetohydrostatic equilibrium from the system of nonlinear MHD equations. Spatial derivatives are discretized using fourth-order centered differences and the solution is advanced in time using a fourth-order Runge-Kutta scheme. It is stabilized by artificial diffusion terms and its parallel design is based on a domain decomposition scheme. A perfectly matched layer (PML) boundary condition \citep{Berenger1994} is applied in order to avoid wave reflections. 

\begin{figure}[!ht] 
 \centering
 \includegraphics[width=8cm]{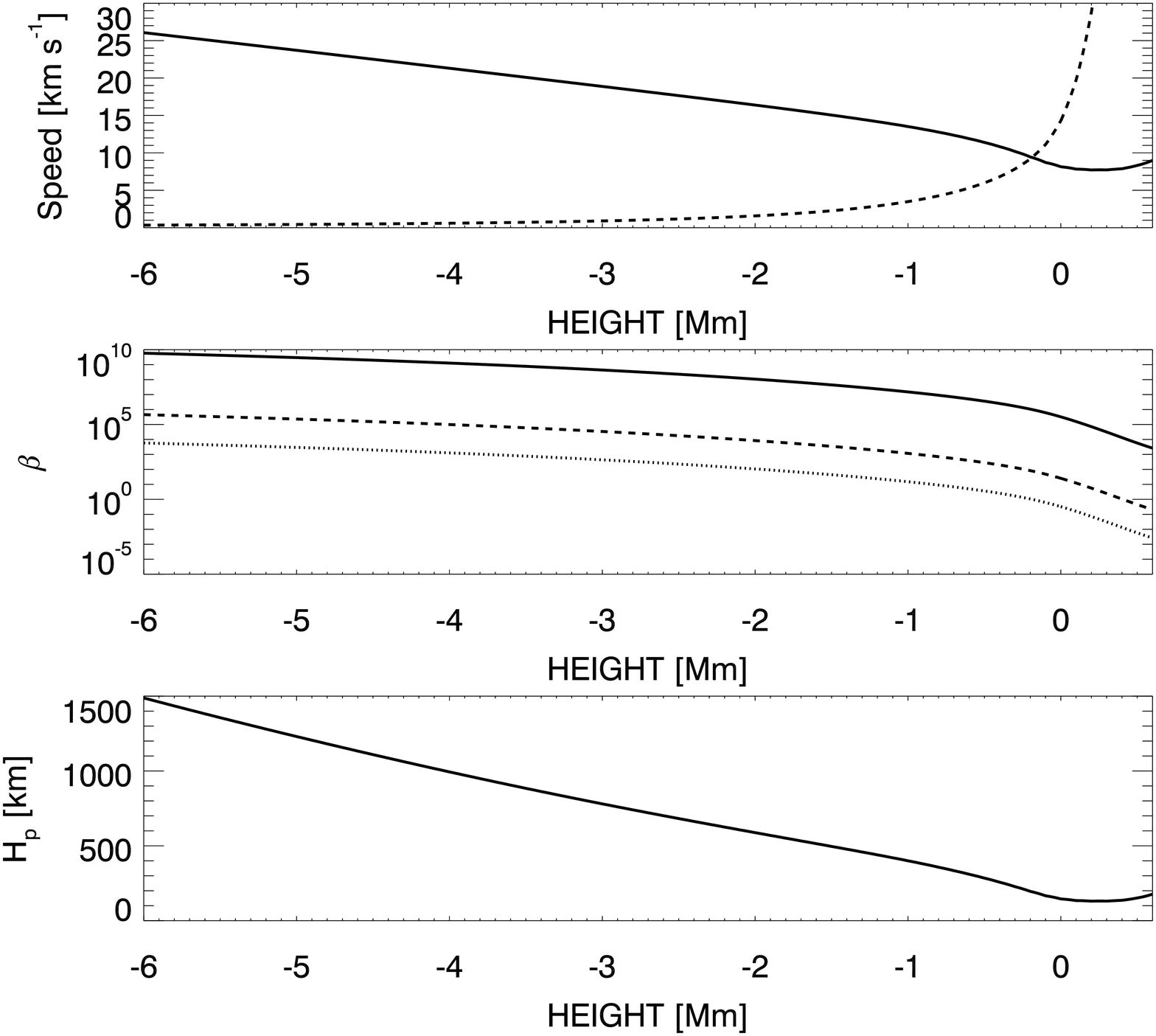}
  \caption{Properties of the flux tubes at the axis. Top panel: sound speed (solid line) and Alfv\'en speed (dashed line) for the 560 km radius tube; middle panel: $\beta$ parameter for the tubes with radius of 170 km (dashed line), 370 km (solid line), and 560 km (dotted line); bottom panel: pressure scale height.}
  \label{fig:tube}
\end{figure}

As a magnetostatic background, we have used flux tube models constructed using the method of \citet{Pneuman+etal1986}, following the routines by \citet{Khomenko+etal2008a}. We consider three flux tube models: the small flux tube has a radius of 170 km and a photospheric magnetic field strength around 1600 G, with slight variations with height and radial distance inside the tube; the medium flux tube has the same magnetic field strength, but a radius of 370 km; and the larger flux tube has also 1600 G photospheric field strength and a radius of 560 km. The external atmosphere outside the flux tube is the quiet Sun model S \citep{Christensen-Dalsgaard+etal1996} stabilized following the method described by \citet{Parchevsky+Kosovichev2007} to avoid convective instability. The radial transition between the magnetic and non-magnetic atmosphere is performed smoothly using a cosine which reduces the magnetic field from its maximum to zero over 100 km in order to avoid numerical problems due to the discontinuity in the magnetic field. For the small flux tube the magnetic field is strictly zero for radial distances higher than 250 km, in the case of the medium tube it vanishes at 450 km, while for the larger tube the magnetic field is zero beyond a radius of 630 km. Figure \ref{fig:tube} shows the characteristic velocities, the $\beta=P_{gas}/P_{mag}$ parameter, where $P_{gas}$ is the gas pressure and $P_{mag}$ the magnetic pressure, and the pressure scale height at the axis of the tubes.

We use a local Cartesian geometry defined by the horizontal coordinates $x$ and $y$ and the vertical coordinate $z$. The computational domain spans from $z=-6$ Mm to $z=0.6$ Mm, where $z=0$ corresponds to the height where the optical depth is unity at a wavelength of 5000 \AA\ in the quiet Sun atmosphere. The horizontal extent of the domain is $x \in [-42.3,32.3]$ Mm and $y \in [-32.5,32.5]$ Mm, with the axis of the vertical flux tube located at $x=0$, $y=0$ Mm. The spatial step is 50 km in the three spatial dimensions. In the vertical direction, PML layers with a thickness of 5 grid cells were set in the top and the bottom boundaries, so the physical domain spans from $z=-5.75$ Mm to $z=0.35$ Mm. In the $x$ direction a 10 points PML was used, while in the $y$ direction we set periodic boundary conditions with no PML.

We aim to study the scattering of an $f-$mode by the flux tube. The vertical velocity of an $f-$mode wave packet which propagates from left to right in the x direction in a horizontally homogeneous atmosphere is described by \citet{Cameron+etal2008} as:

\begin{equation}
\label{eq:fmode}
v_z(x,y,z,t)=Re\sum_k A_k e^{kz}e^{ik(x-x_0)-i\omega_kt}
\end{equation}

\noindent where $Re$ means real part, $A_k$ are complex amplitudes, $x_0$ indicates the initial position of the wave packet, $t$ is time, and $\omega_k=\sqrt{g_0k}$ is the eigenfrequency at wavenumber $k$. This wave packet is uniquely determined by the initial conditions:

\begin{equation}
\label{eq:fmode_t=0_v}
{\bf v}=Re\sum_k(i{\bf \hat{x}}+{\bf \hat{z}}) A_k e^{kz+ik(x-x_0)},
\end{equation}

\begin{equation}
\label{eq:fmode_t=0_p}
p_1=Re\sum_k i A_k \omega_k^{-1}\rho_0 g_0 e^{kz+ik(x-x_0)},
\end{equation}

\begin{equation}
\label{eq:fmode_t=0_rho}
\rho_1=Re\sum_k i A_k \omega_k^{-1}\rho_0H_{\rho}^{-1} e^{kz+ik(x-x_0)},
\end{equation}

\noindent where $p_1$ and $\rho_1$ are the Eulerian perturbations in pressure and density, respectively, $\rho_0$ is the unperturbed density, ${\bf g}=-g_0{\bf \hat{z}}$ is the acceleration due to gravity,  ${\bf \hat{z}}$ is a unit vector pointing upward, and $H_{\rho}$ is the density scale height given by the expression:

\begin{equation}
\label{eq:H_rho}
\frac{1}{H_{\rho}}=\frac{1}{H_p}+\frac{1}{T_0}\frac{dT_0}{dz},
\end{equation}

\noindent where $T_0$ is the background temperature, $H_p=c_s^2/(\gamma g_0)$ is the pressure scale height, and $\gamma$ is the ratio of specific heats, and $c_s$ is the sound speed. Equations (\ref{eq:fmode_t=0_p}) and (\ref{eq:fmode_t=0_rho}) are obtained from the initial displacement vector for an $f-$mode \citep[Equation (19) from][]{Cameron+etal2008} and its relations with the perturbations in density and pressure \citep[\eg, Equations (4) and (5) from][respectively]{Cameron+etal2008}.

Following Eqs. (\ref{eq:fmode_t=0_v})-(\ref{eq:fmode_t=0_rho}) we have introduced an $f-$mode at $t=0$ s located at $x_0=-37.3$ Mm and spanning along the full domain in the y direction. As an initial distribution of $f-$mode amplitudes $A_k$ as a function of $L=kR_{\odot}$, where $k$ is the horizontal wavenumber and $R_{\odot}$ is the solar radius,  we have imposed a Gaussian centered at $L=1000$ and with a half width of $600$. This is an unrealistic distribution, but we are interested in quantifying the absorption and phase shift at each $L$ rather than reproducing the solar spectrum. We have chosen the distribution of $f-$mode amplitudes in order to get enough power in the wavenumbers of interest. Since we took real values for $A_k$, all waves with different $k$ are in phase at the starting position. We have imposed a small amplitude in order to be sure that the simulation remains in the linear regime. The initial location of the wave packet $x_0$ was selected to introduce the wave outside of the outer circumference of the Hankel analysis. The duration of the simulation is $T=180$ min, which corresponds to the total time that the wave packet needs to travel through all the domain in the $x$ direction. The output is saved every minute, producing a set of 181 three-dimensional cubes that provide the temporal evolution of the three components of the velocity, pressure, density, and the three components of the magnetic field for all the computational domain. 

Together with the 3D simulations of the atmosphere with the flux tube, we have also performed a two-dimensional simulation in a quiet Sun atmosphere, without the flux tube being present, but otherwise using exactly the same configuration as in the flux tube model computation. According to the set up of the 3D simulation, all the grid rows in the $x$ direction for a corresponding $y$ are equivalent, except for the presence of the tube, which allows us to use the 2D quiet Sun simulation as a reference to obtain the scattered wave as the difference between both simulations by subtracting the 2D simulation from all the $xz$ planes in the 3D computation.

\section{$f-$mode scattering}
\label{sect:scattering}

\begin{figure}[!ht] 
 \centering
 \includegraphics[width=9cm]{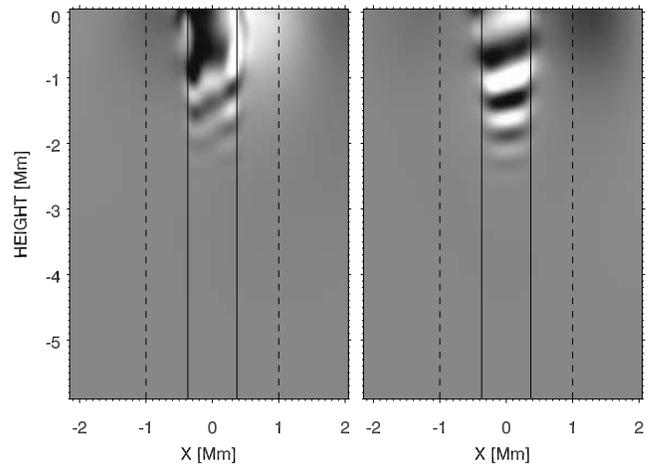}
  \caption{Vertical cuts of the $z$-component of velocity (left panel) and the $x$-component of the velocity (right panel) scaled with factor $\rho_0^{1/2}$ at $t=100$ min in the presence of a 370 km radius flux tube after substracting the quiet Sun simulation. The white-black colors mean positive-negative velocity directions; the range of the grey scale is the same in both panels. Vertical solid lines represent the boundaries of the flux tube. Vertical dashed lines correspond to the outer limits of the jacket modes region (see Figure \ref{fig:flux_jacket}), although they are only visible close to the tube.}
  \label{fig:conversion}
\end{figure}

When the $f-$mode reaches the flux tube different tube waves are excited. These waves propagate upward and downward along the magnetic field lines, extracting energy from the acoustic cavity and producing an effective absorption of the $f-$mode energy. The tube waves are visible in Figure \ref{fig:conversion}, which shows the $z$ velocity (left panel) and $x$ velocity (right panel) scaled with factor $\rho_0^{1/2}$ at $t=100$ min in a region in and around the flux tube. As these waves propagate downward their amplitude drops due to the higher density at deeper layers, while their wavelength also decreases because of the lower Alfv\'en velocity.

\begin{figure}[!ht] 
 \centering
 \includegraphics[width=9cm]{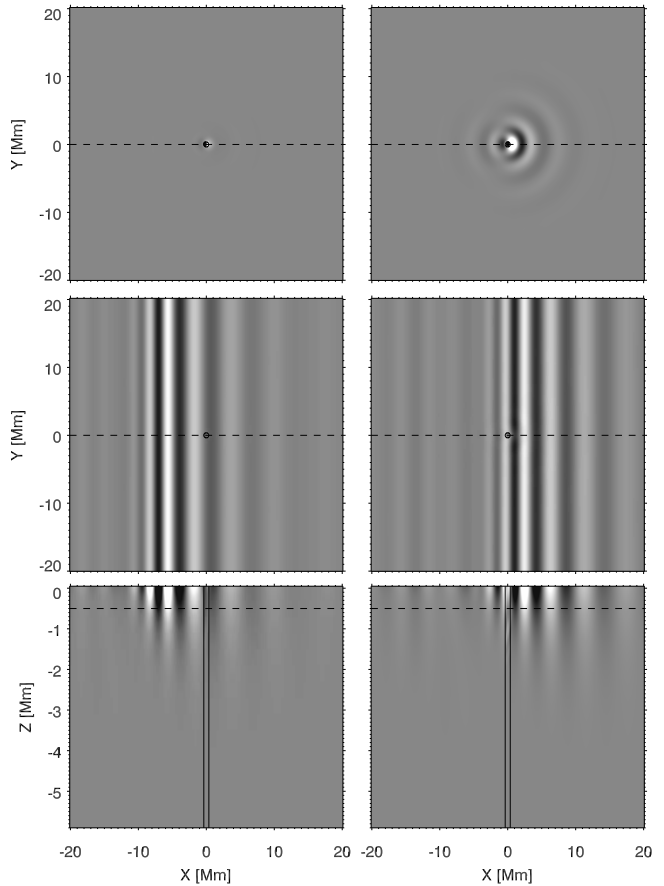}
  \caption{Vertical velocity at $t=80$ min (left panels) and $t=100$ min (right panels) for the simulation with a 370 km radius magnetic flux tube. Top: horizontal cut at $z=-0.5$ Mm of the scattered wave; middle: horizontal cut at $z=-0.5$ Mm of the full wave field; bottom: vertical cut at $y=0$ Mm for the full wave field. The white-black colors mean positive-negative velocity directions; the grey scale in the top panels is 10 times more saturated. In the top and middle panels the circle at $x=0$ Mm and $y=0$ Mm indicates the location of the tube, while the dashed line is the position of the cut shown in bottom panels. In the bottom panels, the vertical solid lines are the boundaries of the tube and the horizontal dashed line is the position of the cuts shown in top and middle panels.  In the full wave field plots the tube modes are hardly visible because their amplitude is small compared to the $f-$mode. See Figure \ref{fig:conversion} for a detalied plot of the tube wave velocities.}
  \label{fig:scattering}
\end{figure}

Figure \ref{fig:scattering} shows the vertical velocity at two different time steps for the simulation with the medium flux tube with $R=370$ km. The left hand column corresponds to $t=80$ min, just before the main part of the wave packet reaches the flux tube, while the right hand column illustrates the simulation at $t=100$ min. The top panels represent the scattered waves in a $xy$ plane at $z=-0.5$ Mm, while the middle and bottom panels show the full wave field in a $xy$ cut at the same height and a $xz$ cut at $y=0$ Mm, respectively. Note that, in order to improve visualization, the color scale is different in the top two panels, which are 10 times more saturated.

The scattered wave is obtained by subtracting the quiet Sun two dimensional simulation from every $xz$ plane along the $y$ direction of the flux tube simulation. In the 170 km radius flux tube the amplitude of the vertical velocity of this wave is 0.020 times the amplitude of the incident $f-$mode, while for the medium tube (with 370 km radius) the ratio between both amplitudes is 0.093. In the case of the big tube (with 560 km radius) the ratio of the amplitudes is 0.235. Since all tubes have the same magnetic field strength, their magnetic flux is proportional to $R^2$, where $R$ is the tube radius. We find that the scattering amplitude approximately scales with the magnetic flux, that is, with $R^2$. 

The oscillations of the waves which travel along the tube are basicaly the generators of the scattered wave, which mainly corresponds to an $f-$mode, although a small amount of power is also scattered into the $p_1-$mode (see Section \ref{sect:mixing}). Since the scatterer is axisymmetric, no scattering can be produced from an azimuthal order $m$ to a different one. In Figure \ref{fig:scattering} it is seen that the scattered wave produced by the 370 km radius tube is composed by a mixture of $m=0$ and $m=\pm1$ waves, dominated by the later ones.

\section{Jacket modes}
\label{sect:jacket}

In addition to the excitation of tube waves and the scattering into different wave modes presented in the previous sections, a scattering object can also generate jacket modes in its surroundings. Jacket modes were first discussed in the context of solar acoustic oscillations by \citet{Bogdan+Cally1995}. They consist of a continuous spectrum of horizontally evanescent wave modes which propagate vertically in the non-magnetic region around the flux tube. The jacket modes are necessary because the vertically evanescent $f$- and $p$-modes alone cannot ensure the continuity of pressure and horizontal velocity of the oscillations across the flux tube boundary, due to the presence of the tube waves that propagate downward at large depth.

The jacket modes obtained in the medium tube simulation can be seen in Figure \ref{fig:conversion} as the downward propagating waves outside the boundaries of the tube with small vertical wavelength, especially in the horizontal velocity (right panel).

The energy extracted by the flux tube from the $f-$mode goes to the tube waves and the jacket modes, which transport the energy upward and downward removing energy from the acoustic cavity. It is interesting to evaluate how the energy is distributed among these modes. In this analysis we will only consider the energy which goes to deeper layers, represented by a negative energy flux, since the proximity of the top boundary hinders obtaining a reliable positive flux into the atmosphere. The wave energy fluxes were calculated following \citet{Bray+Loughhead1974}. The acoustic energy flux is given by the expression:
\begin{equation}
{\bf F_{ac}}=p_1{\bf v},
\label{eq:Fac}
\end{equation}
and magnetic wave energy flux is given by:
\begin{equation}
{\bf F_{mag}}={\bf B_1}\times({\bf v}\times {\bf B_0})/\mu_0.
\label{eq:Fmag}
\end{equation}
 
\noindent In these expressions ${\bf v}$ and ${\bf B_1}$ are the perturbed velocity and magnetic field, respectively, ${\bf B_0}$ is the background magnetic field and $\mu_0$ is the magnetic permeability. 

\begin{figure}[!ht] 
 \centering
 \includegraphics[width=9cm]{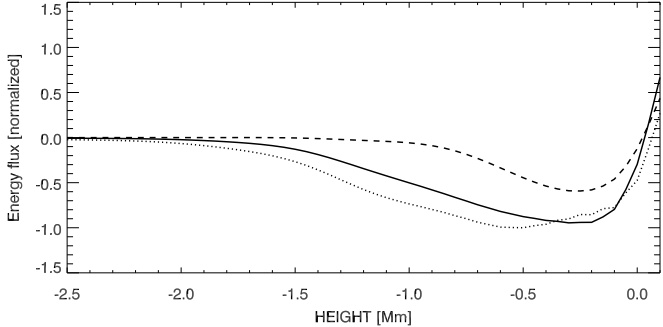}
  \caption{Total vertical wave energy flux inside the magnetic flux tube with 560 km radius (dotted line), the tube with 370 km radius (solid line), and the tube with 170 km radius (dashed line). All cases are normalized to the absolute value of the flux of the 560 km radius tube at $z=-0.5$ Mm.}
  \label{fig:flux_tube}
\end{figure}

In the case of the waves inside the tube, the acoustic flux inside the tube can be obtained as the difference between the acoustic flux in the simulation with the flux tube being present and the quiet Sun simulation. With regards to the magnetic wave flux, it is directly obtained from the application of Equation (\ref{eq:Fmag}) to the region where the magnetic field is different from zero. Adding both fluxes we retrieve the total energy flux carried downward by the tube waves. At each height, we have summed the energy flux correponding to all the points inside the tube for all the time steps between the time that the $f-$mode wave packet reaches the tube and the time that the wave packet leaves it. The result is shown in Figure \ref{fig:flux_tube}, where the variation of the wave energy flux of the tube waves with the height is plotted. Around $z=0$ km the total flux vanishes. Above that height the flux is positive, which means that the energy propagates upward, and below $z=0$ Mm it is negative, showing a minimum around $z=-0.2$ Mm in the case of the two smaller tubes and around $z=-0.5$ Mm for the larger tube. As these waves propagate downward they are damped by the diffusivity, and their energy flux tends to zero at about $z=-2$ Mm in the case of the 370 km and 560 km radius tube and at $z=-1$ Mm for the 170 km radius tube. The downward wave energy flux for the two bigger tubes is similar, being 1.6 times higher than the corresponding to the small tube. 

The jacket modes appear in the non-magnetic region, surrounding the magnetic flux tube. Thus, only the acoustic wave flux contributes to its wave energy. However, some care must be taken to obtain a correct evaluation of its energy flux. Outside the flux tube, the scattered wave field (obtained as the difference between the simulation with the flux tube and the quiet Sun simulation) is composed by the scattered wave (fundamentally an $f-$mode) and the jacket modes. The difference between the acoustic flux in both simulations will correspond to the sum of the scattered wave and the jacket mode fluxes. To isolate the contribution of the jacket modes, we have filtered out the low vertical wavenumbers of the scattered velocity in the surroundings of the flux tube, and we have calculated its acoustic flux using that velocity in Equation \ref{eq:Fac}. The filter selects the waves with vertical wavenumber higher than $3.14$ Mm$^{-1}$ and, thus, only waves with vertical wavelength below $2$ Mm are considered for the jacket modes. At the photosphere the jacket modes have a vertical wavelength around 1 Mm. As they propagate downward, their wavelength is reduced in order to match the decrease of the wavelength of the tube waves due to the reduction of the Alfv\'en speed, showing a 0.45 Mm vertical wavelength at $z=-3$ Mm.

\begin{figure}[!ht] 
 \centering
 \includegraphics[width=9cm]{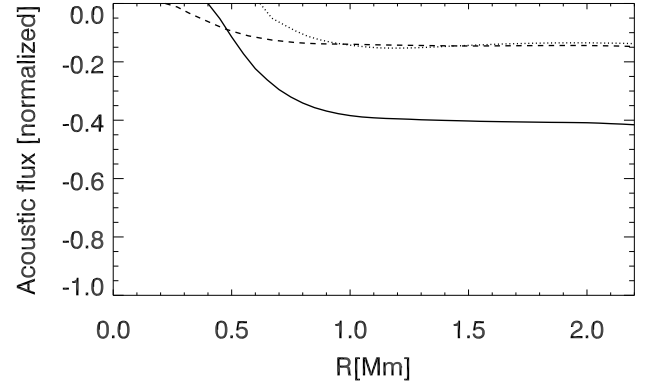}
  \caption{Total vertical acoustic wave energy flux of the jacket modes inside an annular region surrounding the flux tube at $z=-0.2$ Mm. The inner radius of the annulus is given by the radius of the flux tube and the outer radius by the abscissa coordinate. Dotted line: 560 km radius; solid line: 370 km radius; dashed line: 170 km radius. The two later cases are normalized to the energy flux of the tube waves at $z=-0.2$ Mm, while the 560 km case is normalized to the energy flux of the tube waves at $z=-0.5$ Mm.
}
  \label{fig:flux_jacket}
\end{figure}

We have summed the energy flux of the jacket modes in an annular region surrounding the flux tube at the height where the downward tube wave energy flux is maximum, that is, at $z=-0.2$ Mm for the two smaller tubes and at $z=-0.5$ Mm for the larger tube. The annulus is delimited by the radius of the flux tube in the inner part, and the size of the outer radius varies. Figure \ref{fig:flux_jacket} shows the result. In the case of the 370 km radius flux tube (solid line) the energy flux of the jacket modes increases with the outer annulus radius until $R=1$ Mm. For larger radius the energy is constant, which means that at larger distances from the tube there is no energy flux associated with the jacket modes. Thus, the size of the jacket mode region is about $0.6$ Mm around the tube. The energy flux has been normalized to the absolute value of the energy of the waves inside the tube at the same height. We find that the energy carried downward by the jacket modes is approximately $40\%$ of the tube waves energy for the 370 km tube. With regards to the 170 km and 560 km flux tubes, a similar size of the jacket mode region is obtained. However, in these cases the energy transported by the jacket modes is around $16\%$ of the energy extracted by the waves inside the tube.

\section{Hankel analysis}
\label{sect:hankel}

The numerical simulations were also analyzed using the Fourier-Hankel spectral decomposition method, as explained in Section \ref{sect:observations}. The coefficients $A_m(L,\nu)$ and $B_m(L,\nu)$ are evaluated from the vertical velocity at $z=0.2$ Mm. A different sampling in $L$ and $\nu$ was obtained because of the differences in the annular domain and duration of the temporal series used. In this case, the analysis is restricted to the annular region delimited by the polar angles $\theta_{min}=0.00589$ rad and  $\theta_{max}=0.04310$ rad, which correspond to a radial distance of $R_{min}=4.1$ Mm and $R_{max}=30$ Mm, respectively. Thus, we obtain a grid with spacings $\Delta L=2\pi/(\theta_{max}-\theta_{min})=168.8$, while the duration of the simulations $T=180$ min provides a $\Delta\nu=1/T=0.0936$ mHz. The outer radius of the annulus is given by the horizontal size of the computational domain. It was chosen as a compromise between a big enough domain to obtain good sampling in $L$, but not too big to avoid very computationally expensive simulations.   

To determine the absorption coefficient, the power of the ingoing and outgoing Hankel components has been summed across the ridge of the $f-$mode in order to retrieve a better signal-to-noise level. At each given $L$, the power for the ingoing wave is determined as

\begin{equation}
\label{eq:Paverage}
|A_m(L)|^2=\int_{\nu_0(L)-\delta\nu}^{\nu_0(L)+\delta\nu}|A_m(L,\nu)|^2d\nu.
\end{equation}

\noindent The frequency window $\delta\nu$ was set to approximately 0.3 mHz, which corresponds to 3-4 frequency bins. The same average is applied to the outgoing $B_m(L,\nu)$ components.  

The absorption coefficient $\alpha_m(L)$ along the ridge of the $f-$mode is then obtained as

\begin{equation}
\label{eq:alpha}
\alpha_m^{def}(L)=1-|B_m(L)|^2/|A_m(L)|^2,
\end{equation}

\noindent and the phase shift is given by Equation (\ref{eq:phase}), without applying the ensemble sum over magnetic elements (me).

The definition of the absorption coefficient may not correspond to a real dissipation of energy if there is significant mode mixing present.

We have calculated the absorption coefficient of the simulation with the flux tube ($\alpha_m^{FT}(L)$) as well as the quiet Sun reference simulation ($\alpha_m^{QS}(L)$). Although in the latter one the absorption should vanish, since there is not any scattering element, due to numerical reasons (numerical diffusivity and interaction with the top boundary) some absorption appears at high wavenumbers. The absorption coefficient measured directly from the quiet Sun simulation is given by

\begin{equation}
\label{eq:alpha_QS_num}
\alpha_m^{QS}(L)=1-\frac{|B_m^{QS}(L)|^2}{|A_m^{QS}(L)|^2}\sigma (L),
\end{equation}

\noindent where $A_m^{QS}(L)$ and $B_m^{QS}(L)$ represent the ingoing and outgoing power is the quiet Sun simulation, respectively, in the ideal case where there is no numerical damping. The numerical damping which produces a different power in the outgoing waves is included in $\sigma (L)$. In the same way, from the flux tube simulation we measure the absorption coefficient as

\begin{equation}
\label{eq:alpha_FT_num}
\alpha_m^{FT}(L)=1-\frac{|B_m^{FT}(L)|^2}{|A_m^{FT}(L)|^2}\sigma (L).
\end{equation}

We assume that the numerical damping $\sigma (L)$ is the same in both simulations, since they use exactly the same configuration. In the quiet Sun the ingoing power is equal to the outgoing power ($|A_m^{QS}(L)|^2=|B_m^{QS}(L)|^2$). Thus, from Equation (\ref{eq:alpha_QS_num}) we retrieve $\sigma (L)=1-\alpha_m^{QS}(L)$. Taking into account that by definition the real absorption coefficient produced by the tube is

\begin{equation}
\label{eq:alpha_FT}
\alpha_m(L)=1-\frac{|B_m^{FT}(L)|^2}{|A_m^{FT}(L)|^2},
\end{equation}

\noindent including Equation (\ref{eq:alpha_FT_num}) and the expresion for $\sigma (L)$ obtained from Equation (\ref{eq:alpha_QS_num}) in the previous equation, we obtain

\begin{equation}
\label{eq:alpha_corrected}
\alpha_m(L)=\frac{\alpha_{m}^{FT}(L)-\alpha_{m}^{QS}(L)}{1-\alpha_{m}^{QS}(L)}.
\end{equation}

In the following we will discuss the absorption coefficient obtained after applying this correction.

In Section \ref{sect:scattering} we discussed qualitatively the properties of the scattered wave. We have performed the Hankel analysis of that wave by decomposing in Hankel functions the difference in the photospheric vertical velocity at $z=0.2$ Mm between the wave field of the simulation with the flux tube being present and the quiet Sun simulation.  In the simulation with the 170 km radius magnetic flux tube the analysis reveals that the power of the components with azimuthal order $m=\pm 1$ is 1.70 times higher than the power of the axisymmetric components with $m=0$. The power in higher azimuthal orders is negligible. For the medium tube (with radius of 370 km) the power of the $m=\pm 1$ components is only 1.19 times higher than the correponding to $m=0$. This simulation shows a small amount of power in the azimuthal orders $m=\pm2$. On the other hand, in the 560 radius flux tube the dominant azimuthal order of the scattered wave is the axisymmetric $m=0$, whose power is 1.1 times higher than the power of $m=\pm 1$. It shows a significant amount of power in $m=\pm 2$, which is 7.43 times smaller than the power in $m=0$.

\subsection{$f-$mode absorption}
\label{sect:absorption}

\begin{figure}[!ht] 
 \centering
 \includegraphics[width=9cm]{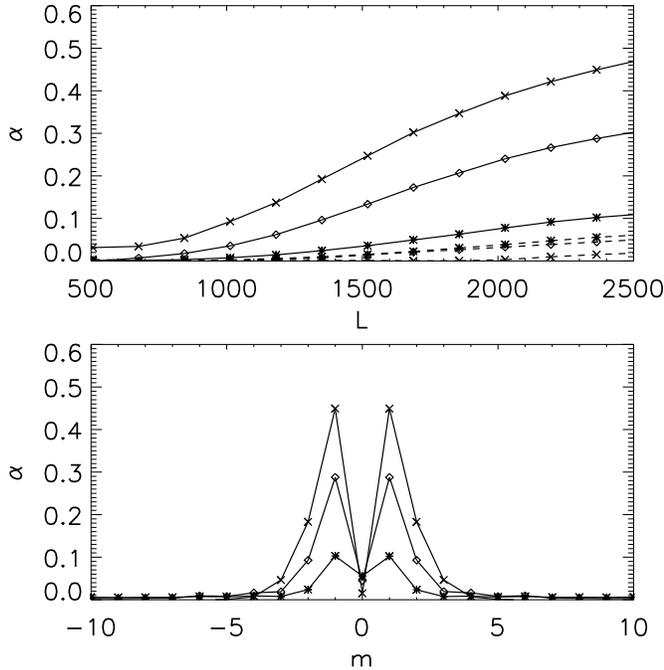}
  \caption{Top panel: Variation of the absorption coefficient with L for the azimuthal orders $m=0$ (dashed line) and $m=1$ (solid line). Bottom panel: Variation of the absorption coefficient with m at $L=2364$. In both panels asterisks correspond to the simulation with a flux tube with a 170 km radius, diamonds to the flux tube with a 370 km radius, and crosses to the flux tube with 560 radius.}
  \label{fig:alpha}
\end{figure}

The top panel of Figure \ref{fig:alpha} shows the absorption coefficient as a function of $L$ for the two lowest azimuthal orders determined for each of the magnetic flux tubes. On the one hand, in the case of the flux tube with a radius of 170 km (asterisks) the highest absorption is retrieved for the azimuthal order $m=\pm1$.  It reaches a value above $0.1$ at $L=2500$. The azimuthal order $m=0$ also shows a significant amount, presenting an absorption higher than half of the corresponding to  $m=\pm1$. For the rest of the azimuthal orders the absorption is very low, although high wavenumbers show some absorption at $m=\pm2$. The variation of the absorption with the azimuthal order is clearly shown in the bottom panel of Figure \ref{fig:alpha} for $L=2364$. It exhibits a perfect symmetry around $m=0$, with clear peaks at $m=\pm1$ and dropping to zero at higher $m$.

On the other hand, for the medium flux tube with a radius of 370 km (diamonds) the measured absorption coefficient is generally higher. The highest absorption is also obtained for $|m|=1$, but in this case its value is around $0.3$ at $L=2500$. Below $L=1500$, the azimuthal orders $m=0$ and $m=2$ (not shown in the figure) produce a similar absorption, but above that $L$ they split up. The absorption of the later one keeps increasing with $L$ and its $\alpha$ is around 0.1 at $L=2500$, while the increase of the absorption in $m=0$ seems to be slower and it shows a maximum absorption around $0.05$ at $L=2500$. The variation of $\alpha$ with $m$ (bottom panel of Figure \ref{fig:alpha}) also shows symmetry around $m=0$.

With regards to the 560 km radius flux tube, the highest absorption is also retrieved for $m=1$, which presents an $\alpha=0.47$ at around $L=2500$. At high $L$ values the absorption coefficient shows significant absorption at $m=2$ and even at $m=3$, opposite to the $m=0$ case, which presents a very low $\alpha$ coefficient.

Several conclusions can be extracted from the comparison of the absorption coefficient measured for different flux tubes. Firstly, the variation with $L$ shows a similar tendency for all flux tubes. It increases almost linearly with $L$ for all the azimuthal orders. Secondly, the absorption is subject to the magnetic flux of the scattering element. A higher magnetic flux produces higher absorption, although its distribution in wavenumber and azimuthal order depends on the radius of the scatterer. Despite the fact that the big flux tube has almost 11 times higher magnetic flux than the smaller one, its absorption coefficient in $m=1$ is far from being infered as 11 times higher than the $m=1$ absorption of the small tube, since the ratio between the absorption of both tubes varies with $L$. At $L=1013$ the $m=1$ absorption of the big flux tube is 16.5 times higher than the corresponding to the small tube, while at $L=2532$ the ratio is just $4.3$. Finally, the relation between the absorption at different azimuthal orders is different. As can be seen in the bottom panel of Figure \ref{fig:alpha}, the higher the radius of the tube the higher absorption coefficient in all azimuthal orders except $m=0$. Surprisingly, $\alpha$ in $m=0$ decreases with increasing radius, despite the higher power of the scattered wave in $m=0$ for the larger radius tubes discussed in the previous section.

\subsection{Phase shifts}
\label{sect:phase}

We are interested in the difference between the ingoing and outgoing phases produced by the scattering element. According to \citet{Braun+etal1992}, for a reliable determination of the phase shift between the incoming and outgoing waves it is necessary that the observations last long enough so that the wave packet can travel a distance comparable to the annulus diameter. The temporal duration of our simulations ($T=180$ min) was chosen in order to satisfy this condition. The phase shift was evaluated following Equation (\ref{eq:phase}). The values discussed in this section correspond to the difference between the simulation with the flux tube and the quiet Sun reference simulation.

\begin{figure}[!ht] 
 \centering
 \includegraphics[width=9cm]{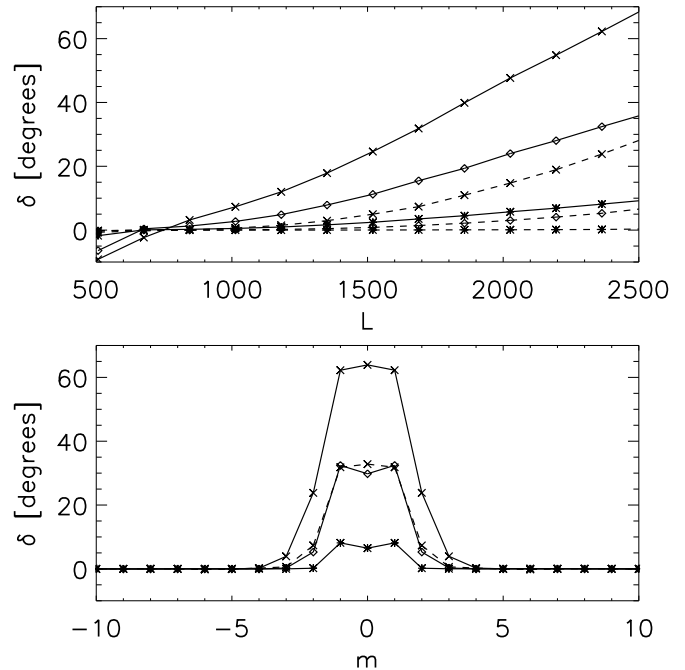}
  \caption{Top panel: Variation of the phase shift with L for the azimuthal orders $m=1$ (solid line) and $m=2$ (dashed line). Bottom panel: Variation of the phase shift with m at $L=2364$. In both panels asterisks correspond to the simulation with a flux tube with a 170 km radius, diamonds to the flux tube with a 370 km radius, and crosses to the flux tube with a 560 radius. The dashed line in the bottom panel corresponds to the 560 km radius flux tube at $L=1765$.}
  \label{fig:phase}
\end{figure}

The top panel of Figure \ref{fig:phase} shows the variation of the phase shift with $L$ for the azimuthal orders $m=1,2$. Starting with the smaller tube (asterisks), we find that the phase shift increases from $\delta_{m=1}=0^o$ at $L=800$ to $\delta_{m=1}\approx10^o$ at $L=2500$. The variation of the azimuthal order $m=0$ with $L$ (not shown in the plot) is very similar to $m=1$, showing an almost linear increase with an slightly lower phase shift. Below $L=800$ the phase shifts of both azimuthal orders are around 0. The azimuthal order $m=2$ shows a much smaller phase shift, and it almost vanishes for all $L$ values. As in the case of the absortion coefficient, significant phase shifts are only obtained for $m=0$ and $m=\pm1$ azimuthal orders (Figure \ref{fig:phase}, bottom panel).

The behavior of the phase shift produced by the medium flux tube (diamonds) is similar to the smaller one, but showing a much higher value. In this case, the phase shift is also around $0^o$ below $L=500$ and it increases with $L$ until reach $\delta\approx 35^o$ for both $m=0$ and $m=1$ azimuthal orders. As in the case of the small flux tube, the phase shift of the $m=1$ azimuthal order is a bit higher, and this difference increases with $L$. The phase shift corresponding to $m=2$ is much lower, since it only reaches $\delta_{m=2}\approx 7^o$.

Finally, the azimuthal order $m=1$ of the larger tube (crosses) shows an increasing phase shift which reaches almost $\delta\approx 70^o$ at $L=2500$. In this case, the phase shift in $m=0$ is slightly higher than for $m=1$, as shown in the bottom panel of Figure \ref{fig:phase}. The phase shift produced in $m=2$ is more significant than in the other tubes, since it is around $\delta\approx 28^o$ at $L=2500$.

A few points deserve special attention. Although the absorption coefficient of $m=0$ decreases with the radius of the tube (Figure \ref{fig:alpha}), it is interesting to note that for the phase shift the azimuthal orders has different relevance. In the bottom panels of Figure \ref{fig:phase} it is clearly seen that the phase shift produced in $m=0$ has a significant role. In fact, the $m=0$ phase shift is very close to the one obtained for $m=1$, and even higher in the case of the larger tube. In all tubes the phase shift shows a similar tendency, being significant for $m=1$ and $m=0$ and small for $m=2$, despite the clear differences which are present in the absorption coefficient. However, the larger radius (and higher magnetic flux) of the bigger tubes does affect the value obtained for the phase shift. At $L=844$ the phase shift obtained for $m=1$ in the big flux tube is 13 times higher than the corresponding to the small flux tube. This difference decreases with $L$, and at $L=2532$ the ratio between the phase shift of both tubes is around $7$. Note that the magnetic flux of the big tube is around 11 times higher than that of the small tube. A similar behavior is obtained for the azimuthal order $m=0$, and also including the medium tube in the comparison. The azimuthal order $m=2$ shows a different pattern, since the ratio of the phase shift between a bigger and a smaller radius tubes increases with $L$.

\subsection{Mode mixing}
\label{sect:mixing}

The absorption coefficient gives us a measure of the power lost by a certain mode, which corresponds to a frequency, wavenumber and azimuthal order. However, it does not necessarily mean that part of its energy has suffered a real absorption. The scattered wave may be a different wave mode, with different degree $L$ and radial order $n$ or azimuthal order $m$. For structures that are stationary in comparison to the typical wave period it is assumed that the outgoing wave must has the same frequency of the incident wave. In our simulations this condition is strictly satisfied, since we are using a magnetohydrostatic model. The change of the incident mode $n$ to a different scattered mode $n'$ at a fixed frequency produced by a magnetic element is commonly called mode mixing. In addition, since these flux tube models are axisymmetric, no scattering can be produced from an azimuthal order $m$ to a different order $m'$

We have tried to measure the mode mixing produced by the three magnetic flux tube models. Since we are introducing as an initial condition the propagation of an $f-$mode, we can only estimate the scattering produced from this incident $f-$mode to higher order modes. However, from these simulations we have only retrieved a significant amount of power in the $p_1$ ridge, so we have evaluated the scattering from the $f-$mode to the $p_1$. We have defined the following quantity

\begin{equation}
\label{eq:alpha_fp}
\alpha_{f-p1}(\nu)=\frac{|B_{p1}^{QS}(\nu)|^2-|B_{p1}^{FT}(\nu)|^2}{|B_f^{QS}(\nu)|^2},
\end{equation}

\noindent where $|B_{p1}^{QS}(\nu)|^2$ is the power in the $p_1$ ridge for the outgoing component of the quiet Sun simulation, $|B_{p1}^{FT}(\nu)|^2$ is the power in the $p_1$ ridge for the outgoing component of the simulation with the flux tube, and $|B_f^{QS}(\nu)|^2$ is the power in the $f-$mode ridge for the outgoing component of the quiet Sun simulation. All these expressions correspond to the sum of the power for the azimuthal orders $m=-1,0,1$ and the degrees $L$ over which the ridge of corresponding mode spans. The coefficient $\alpha_{f-p1}$ represents the ratio between the power in the $p_1$-mode generated by the flux tube and the power in the $f-$mode in the case without the tube being present at the same frequency. A negative value means that there has been emission to the $p_1-$mode. In the numerator we have introduced the difference in power in the $p_1$-mode between the quiet Sun and the flux tube simulations in order to correct from the small amount of energy which appears in this mode from the initial condition. We have introduced as initial condition the analytical expression of a propagating $f-$mode. However, due to the limited size of our computational domain and the presence of PML layers the solution of an $f-$mode in our domain is slightly different, and a very small amount of power goes to the $p_1$ ridge.

\begin{figure}[!ht] 
 \centering
 \includegraphics[width=9cm]{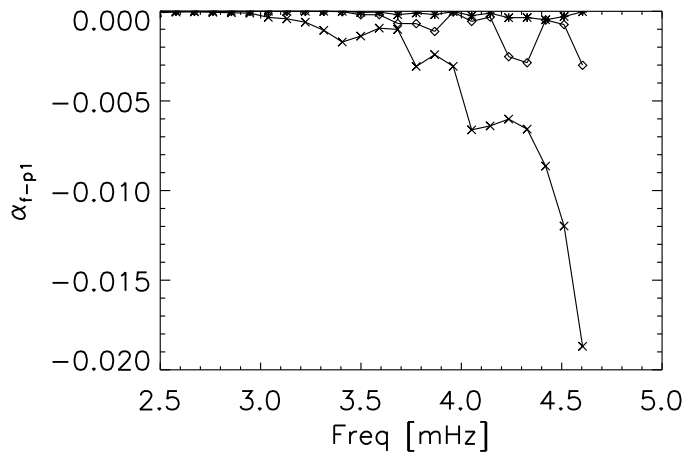}
  \caption{Variation of the coefficient $\alpha_{f-p1}$ (Equation (\ref{eq:alpha_fp})) with frequency. Crosses: 560 km radius flux tube; diamonds: 370 km radius flux tube; asterisks: 170 km radius flux tube.}
  \label{fig:alpha_fp}
\end{figure}

The $\alpha_{f-p1}$ coefficient obtained for all simulations is plotted in Figure \ref{fig:alpha_fp}. No measurable amount of power was found in the $p_1$ ridge below 2.6 mHz and above 4.6 mHz. On the other and, the low power obtained between these frequencies makes it hard to retrieve a reliable measure of the mode mixing. For the large flux tube we find that the power scattered to the $p_1$ mode is below $2\times10^{-2}$ times the power of the incident $f-$mode. The coefficient $\alpha_{f-p1}$ seems to decrease with the frequency, although the result is very noisy. In the case of the medium tube, $\alpha_{f-p1}$ is lower than $3\times10^{-3}$. The smaller tube show a much lower mode mixing, and its $\alpha_{f-p1}$ decreases from $0$ at 3.4 mHz to $-5\times10^{-4}$ at 4.4 mHz.

\subsection{Comparison with observations}

Figure \ref{fig:phase_obs_solo} represents the phase shift obtained from the observations. As shown by the bottom panel, it has a broad distribution in azimuthal order, exhibiting a significant phase shift between $m=-6$ and $m=5$, with a peak at $m=0$. The variation of the phase shift with $m$ greatly differs from the one measured for the simulation, where in the two smaller radius cases we found that the phase shift is concentrated in the azimuthal orders $m=-1,0,1$ with symmetric distribution, peaking at $m=-1$ and $m=1$, while the larger tube shows some significant phase shift in $m=2$ (Figure \ref{fig:phase}).

\begin{figure*}[!ht] 
 \centering
 \includegraphics[width=18cm]{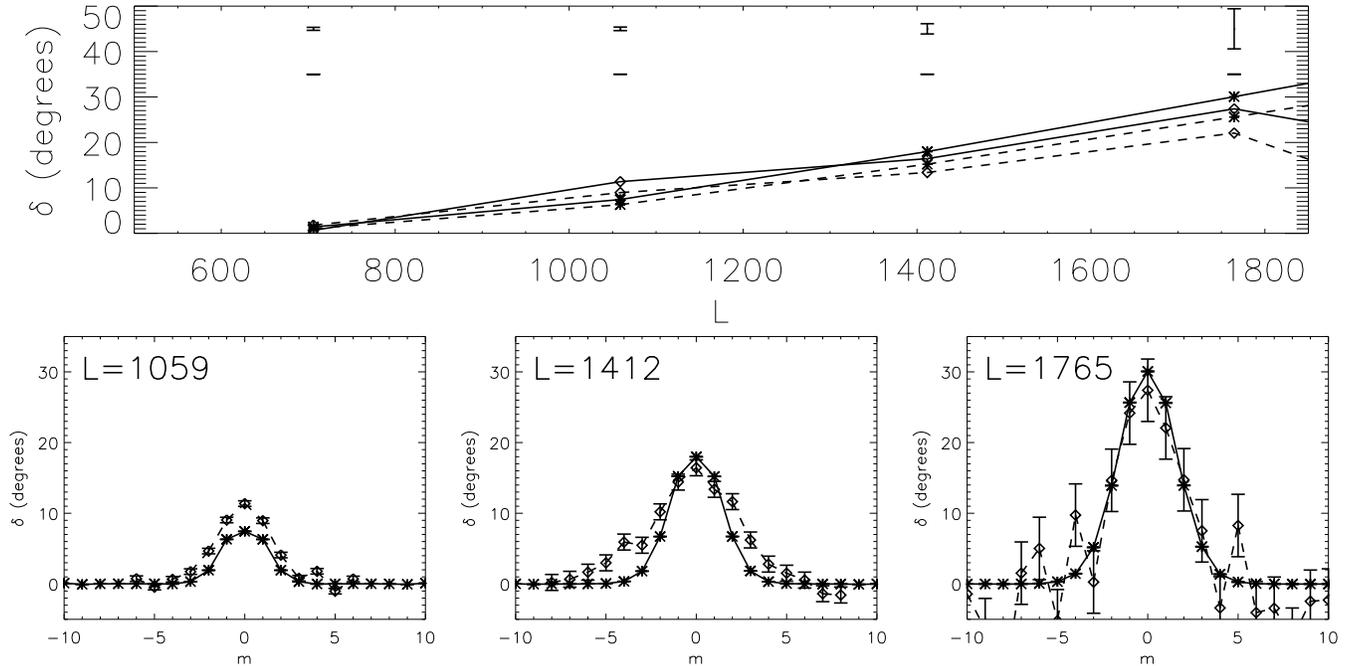}
  \caption{Top panel: Variation of the phase shift with L for the azimuthal orders $m=0$ (solid line) and $m=1$ (dashed line). The bars above each $L$ with data indicate the errors, the bottom one corresponds to the simulation and the top one is the error from the observations. Bottom panels: Variation of the phase shift with m at different $L$; from left to right $L=1056$, $L=1412$, and $L=1765$. In all panels asterisks correspond to the simulation with a flux tube with a 560 km radius after averaging and diamonds to the observations.}
  \label{fig:phase_obs}

\vspace{0.5mm}
\end{figure*}

We consider a scenario which could explain this mismatch. We have tried to mimic the error introduced in the observational Hankel decomposition by a displacement in the position of the annulus with respect to the magnetic element. This spatial translation can strongly affect the incoming and outgoing power retrieved from the Hankel decomposition. As an example, consider a point source of waves which generates a concentric pattern of wavefronts that propagates radially outwards, like the one created by a pebble dropped into a pond. When the origin of the coordinate axis is located at the position where the pebble hits the water, the Fourier-Hankel analysis of the wave will reveal a large $B_0$ (outgoing axisymmetric wave) and zero $A_0$ (no incoming power). However, at different positions the power will appear in different azimuthal orders. As stated above, the observed phase shift is retrieved from 3400 locations of small magnetic elements. If the resolution limitations and the movement of the magnetic element during the 4 hours of observation produce a slight shift in the determination of the center of the magnetic element, the average of the phase shift obtained from the 3400 elements, all of them with different displacement, will produce a broadening of the distribution of the phase shift with the azimuthal order. We have imitated this limitation by averaging the phase shift retrieved from 3400 realizations of the Hankel analysis in the simulation with the 560 km radius tube, with the center of the annulus shifted randomly around the axis of the tube in a Gaussian distribution with FWHM equal to that obtained from an estimation of the Point Spread Function for MDI high-resolution data, which corresponds to 1.14 Mm \citep{Tarbell+etal1997}. In this analysis we have used the same annular region used in the observations in order to obtain the same sampling in $L$.

Figure \ref{fig:phase_obs} shows the phase shift obtained from the observations (diamonds) and the corresponding to the simulations after applying the average (asterisks). The top panel shows that both the simulated and observational phase shifts increase with $L$ for the two azimuthal orders plotted. At $L=1059$ the observational phase shift is higher than the simulated one, but the rest of the $L$ show good agreement. The error in the simulations is obtained using the same estimation as in the observations, although it is very small due to the symmetry retrieved between the positive and negative azimuthal orders. 

In the bottom panels we can see how the simulations resemble the observational broad distribution with $m$. As higher $L$ is considered, significant phase shifts are obtained at higher azimuthal orders. At $L=1412$, in the simulations the phase shift drop to zero at $m=\pm5$, while the observations present a slightly broader distribution, with higher phase shifts at higher $m$. The dashed line in the bottom panel of Figure \ref{fig:phase} shows the phase shift obtained for the 560 km radius tube at $L=1765$, the same $L$ value of the right bottom panel of Figure \ref{fig:phase_obs}. Note the broadening produced by the shifted average of the Hankel analysis. Interestingly, it also produces the effect of generating a clear phase shift peak at $m=0$, which differs from the original data, where the azimuthal orders $m=0$ and $m=1$ shows approximately the same shift.

\section{Discussion and conclusions}
\label{sect:conclusions}

We have presented the analysis of the scattering produced by magnetic flux tube models using 3D numerical simulations. Previous attemps to model this phenomenon \citep[\eg,][]{Gordovskyy+Jain2007, Jain+etal2009,Hanasoge+etal2008, Hindman+Jain2011} have faced the problem by means of analytical treatments. These types of studies are the first steps toward the comprenhension of this issue and they provide a valuable heritage to understand the wave interaction with magnetic media and confront it with the forthcoming observations and modeling. However, the simplifications needed to carry out their development restrict their results to some idealized cases. From this scope, the use of numerical simulations emerges naturally as the next step to address these questions in more general situations. 

In these simulations we have propagated an $f-$mode through a model S atmosphere \citep{Christensen-Dalsgaard+etal1996} stabilized against convective instabilities embedded with a flux tube model. In order to compare how some properties of the tubes affect the scattering three realizations were performed, using different flux tube models. All tubes have the same peak magnetic field strength, but they differ in the radius and, thus, in the magnetic flux. 

Our simulations show that the interaction of an $f-$mode with a flux tube excites tube waves. These waves propagate along the magnetic field lines and produce a real absorption of the incident energy, since it is extracted from the acoustic cavity. The oscillations of the tube produced by these waves generate a scattered wave. It is composed of a mixture of axisymmetric ($m=0$) and dipolar ($m=\pm1$) modes, whose distribution in frequency and azimuthal order depends on the radius of the flux tube. For thin flux tubes, the $m=\pm1$ dipolar oscillation dominates the tube wave, while axisymmetric oscillations ($m=0$) become important for larger tubes. This result agrees with those previously obtained by \citet{Daiffallah+etal2011}. 

We have quantified the absorption coefficient and phase shift produced by the three magnetic flux tube models. Based on the results discussed in the previous section, we draw the following conclusions. Firstly, the absorption increases with wavenumber (frequency) for all azimuthal orders and tube models. Secondly, the amount of absorption in general increases with the magnetic flux of the tube, although this increase depends significantly on the wavenumber and azimuthal order. Thirdly, the distribution of the absorption in azimuthal order depends on the radius of the tube. In all models the peak absorption is obtained for $m=1$. However, in the tube with 170 km radius it is followed by $m=0$, with a weak absorption in $m=2$, while in the 370 km radius tube the absorption in $m=2$ is stronger than the corresponding to $m=0$, and in the case of the 560 km radius tube the absorption in $m=0$ is especially low. The absorption of the axisymmetric $m=0$ order decreases with the radius.

It is noticeable that the different behavior that the phase shift shows regarding the second and third points of the previous paragraph. From the simulations with the tubes we find a similar distribution in the phase shift produced in different azimuthal orders, although it seems to approximately scale with the magnetic flux, with some dependence on the $L$ value and the azimuthal order. In this way, $m=0$ and $|m|=1$ show a very similar phase shift, the later slightly higher in the two smaller tubes and the opposite in the larger tube, while the phase shift produced in $m=2$ is very small, except for the 560 km radius tube.    

In this work we are not only interested in modeling the scattering process, but also in applying this knowledge to interpret observations. A deeper understanding of the wave interaction with small magnetic scatterers can yield a basis to infer the properties of the scattering elements, even at scales smaller than the observational resolution. We have compared the numerical results for the phase shift with observations of an ensemble averaging of thousands of small magnetic elements. In order to perform an equivalent comparison, the phase shift obtained from 3400 realizations of the Hankel analysis of the simulations with the 560 km radius tube with a small shift in the position of the annulus was also averaged.

The phase shift produced by our larger tube model after averaging shows a good qualitative agreement with the observed phase shift. Since the phase shift scales with the magnetic flux of the scattering element, we may consider that the phase shift of the observed elements could be produced by fluxtubes with magnetic flux around the corresponding to tubes with 560 km radius and 1600 G. The current work suggests one possible solution for the properties of the tube model, although other combinations of radius and magnetic field strength might also work. This kind of measurement seems to be a promising method to infer the characteristics of small magnetic networks elements. However, some caution must be considered in their interpretation. Some of the observed magnetic elements used in this study show strong asymmetries (see Figure \ref{fig:magnetogram}). When the scattering element is not axisymmetric the scattering is not restricted to occur from an azimuthal order $m$ to the same order $m$, but the scatterred wave can correspond to a different order $m'$. These nonsymmetric elements could contribute to the broadening of the distribution of phase shift with $m$. On the other hand, in the observed magnetograms it is hard to find enough completely isolated magnetic elements. From the 3400 elements used in the analysis, many of them present other small magnetic features inside the 16.5 Mm annulus in which the Hankel decomposition was performed. The observational analysis might be contaminated by the scattering produced by these other elements. 

In the comparison between the observations and the simulations, we have assumed that the observational shift in the center of the Hankel analysis is restricted to a Gaussian with the FWHM of the PSF from high-resolution MDI. If the proper motions of the flux tube have larger extension, the results could be affected.  However, the tubes are presumably moving at the time scale of the granulation, which is not so different from the time scale of the wave period. This could mean that the approximation of a stationary tube is not so good, and limits the capacity of this work to address this issue. It would be interesting to extend the analysis to moving flux tubes in the future.

These results provide a warning to be cautious when interpreting ensemble averages of observational data. In observations like those presented in this paper, where the contribution of an individual magnetic feature is too low to get a reliable measurement, the average of several cases is a compulsory procedure to obtain a strong enough signal. However, the individual and unique characteristics of each element, together with the limitations to perform the analysis using exactly the same configuration, can lead to a result which may point to conclusions that do not reflect the real observed structure. In the particular case studied in this work, from the observational broad distribution of the phase shift with $m$, one could assume that a big magnetic feature is necessary to produce that dependence with the azimuthal order. The analysis of the averaged simulated flux tube reveals that a similar measure of the phase shift dependence can be retrieved from a very different magnetic element, making difficult infering an irrefutable conclusion about the nature of the observed elements.

\acknowledgements   We thank Tom Duvall for providing observational data and for his helpful comments to an early version of this paper, and the referee for his/her valuable suggestions. This research has been funded by NASA through projects NNH09CE43C, NNH09CF68C, NNH09CE41C, and NNH07CD25C. This work used the Extreme Science and Engineering Discovery Environment (XSEDE), which is supported by National Science Foundation grant number OCI-1053575, and NASA's Pleiades supercomputer at Ames Reasearch Center, together with testing at LaPalma supercomputer at Centro de Astrof\'{\i}sica de La Palma and on MareNostrum supercomputer at
Barcelona Supercomputing Center (the nodes of Spanish National
Supercomputing Center).

\aareferences

\end{document}